PAPER • OPEN ACCESS

# The Kitaev honeycomb model on surfaces of genus $g \geq 2$

PAPER

# The Kitaev honeycomb model on surfaces of genus $g \geqslant 2$

John Brennan[1] and Jiří Vala[1,2]

[1] Department of Theoretical Physics, Maynooth University, Maynooth, County Kildare, Ireland
[2] School of Theoretical Physics, Dublin Institute for Advanced Studies, 10 Burlington Road, Dublin, Ireland

E-mail: Jiri.Vala@mu.ie

**Keywords:** topological phase, topological quantum field theory, topological degeneracy, lattice defect

## Abstract

We present a construction of the Kitaev honeycomb lattice model on an arbitrary higher genus surface. We first generalize the exact solution of the model based on the Jordan–Wigner fermionization to a surface with genus $g = 2$, and then use this as a basic module to extend the solution to lattices of arbitrary genus. We demonstrate our method by calculating the ground states of the model in both the Abelian doubled $\mathbb{Z}_2$ phase and the non-Abelian Ising topological phase on lattices with the genus up to $g = 6$. We verify the expected ground state degeneracy of the system in both topological phases and further illuminate the role of fermionic parity in the Abelian phase.

## 1. Introduction

The Kitaev honeycomb model is an example of an exactly solvable two-dimensional model that exhibits both Abelian and non-Abelian topological phases [1]. The Abelian phase, which is also known as the toric code [2], provides a realization of a topological quantum field theory known as doubled-$\mathbb{Z}_2$ theory. The non-Abelian phase is effectively described by the Ising topological quantum field theory [3]. The main attribute of topological field theories is a dependence of the dimension of the relevant Hilbert space on the topology of the underlying manifold on which these theories are realized. For example the doubled-$\mathbb{Z}_2$ theory is represented in the two-dimensional toric code by a non-degenerate ground state on a genus 0 surface like an infinite plane or a sphere, and a four-fold degenerate ground state on a genus 1 surface like a torus. Similarly, the Ising topological field theory is linked to a three-fold degenerate ground state of the honeycomb lattice model in its non-Abelian phase on a torus. However, the square lattice of the toric code and the honeycomb lattice of the Kitaev model permit realizations of only these two surface topologies as the Euler characteristics of both lattices are zero.

We extend the solution of the Kitaev honeycomb model to closed surfaces of genus greater than one. We will rely on the exact solution of the model based on Jordan–Wigner fermionization [4–6]. This solution allows us to factorize the model into a fermionic superconductor on a topological doubled-$\mathbb{Z}_2$ square lattice background or vacuum state. In order to generalize this to higher genus surfaces, we introduce a lattice which can be realized on such surfaces and accordingly adjust any definitions and relations to this context. We will then demonstrate the generalized solution on a number of different surfaces of genus greater than 1 by calculating the ground state degeneracy of the model in both the Abelian and non-Abelian phases. In this context, we also investigate additional features of these topological states that are intrinsic to their lattice realizations.

A natural framework for our investigation of two-dimensional lattice models whose topological phases effectively realize certain topological quantum field theories is the axiomatic definition of these theories. $n$-dimensional topological quantum field theory is defined as a functor from a category of $n$-cobordisms to a category of vector spaces [7, 8]

$$F\colon n\text{-Cob} \to \text{Vect}. \tag{1}$$

To specify a category, we have to identify its objects and morphisms (or arrows) between them. Specifically, objects of the category of $n$-cobordisms $n$-Cob are closed oriented $(n - 1)$-dimensional manifolds. These form boundaries of an oriented $n$-dimensional manifold which then constitutes a morphism between the objects of the category $n$-Cob. Topological quantum field theory is a rule that associates with $(n - 1)$-manifolds finite-dimensional vector spaces and with each $n$-manifold a map between these vector spaces. Furthermore, this rule is subject to certain axioms which for example ensure that vector spaces originating from topologically equivalent manifolds are isomorphic and that the disjoint union of $(n - 1)$-manifold carries over to a tensor product between vector spaces. The functor satisfying these axioms is called modular and the underlying categories are called monoidal. We point out that realizing a topological phase of a physical system on a closed oriented surface of some genus represents a realization of an important part of this functor. Specifically it assigns to the surface (2-manifold) a vector space spanned by the ground states of the relevant physical system.

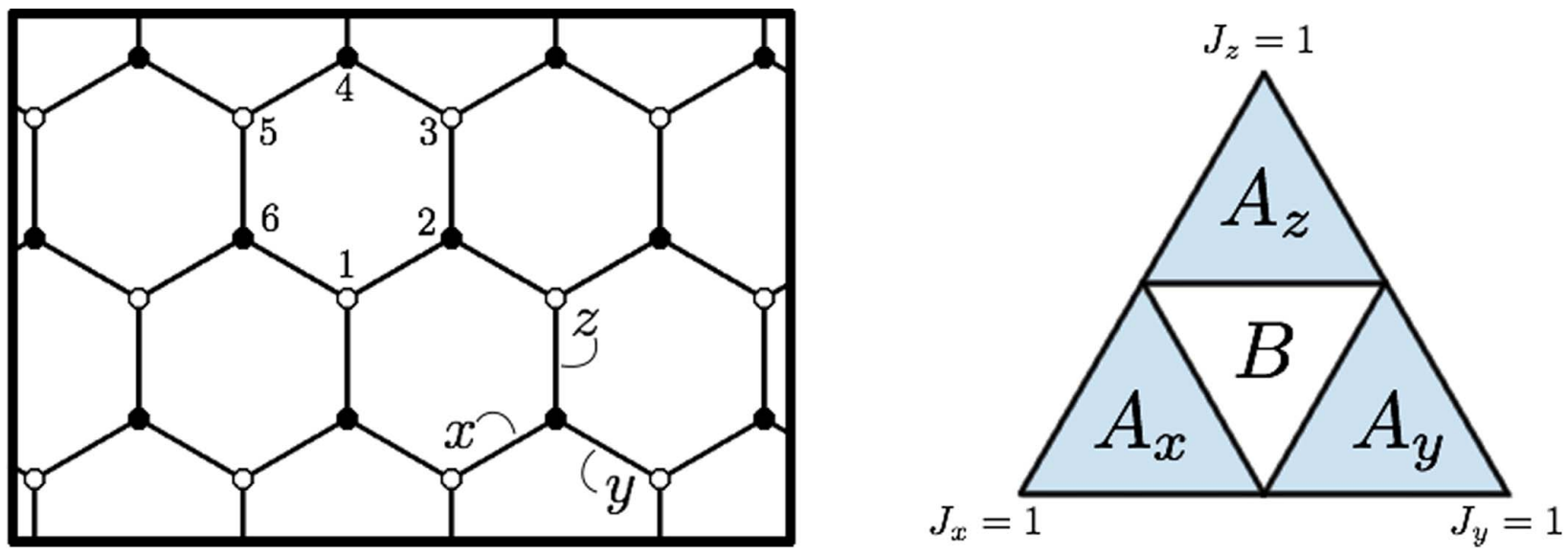

**Figure 1.** The Kitaev honeycomb lattice model and its phase diagram. As shown on the left, any given link has one of three possible orientations, $x$, $y$ or $z$ and we number the vertices of a plaquette 1–6. The phase diagram can be thought of as the convex hull of the three points $(J_x, J_y, J_z) = (1, 0, 0), (0, 1, 0), (0, 0, 1)$.

We first give a concise overview of the model and its effective spin/hardcore-boson representation on a square lattice in section 2. A realization of lattices on higher genus surfaces is introduced in section 3, which is then followed by the implementation of the model on these lattices and its solution using the Jordan–Wigner fermionization in its effective spin/hardcore-boson representation in section 4. The last two sections describe the calculation of the ground state in section 5 and evaluation of the ground state degeneracy of the model on surfaces with the genus 2–6 in section 6.

## 2. The model

The Kitaev model [1] is a honeycomb lattice with a spin $\frac{1}{2}$ particle attached to each vertex. Each spin interacts only with its nearest neighbors via an interaction term that depends on the orientation of the link ($x$, $y$ or $z$) connecting them. Explicitly, if $i$ and $j$ label neighboring vertices connected by a link of orientation $\alpha$, these spins interact via a term of the form $J_\alpha \sigma_i^\alpha \sigma_j^\alpha$. Here $J_\alpha$ is a constant determining the strength of interactions along links of orientation $\alpha$. The model's Hamiltonian is the sum of all these interactions:

$$H_0 = -J_x \sum_{x\text{-links}} \sigma_i^x \sigma_j^x - J_y \sum_{y\text{-links}} \sigma_i^y \sigma_j^y - J_z \sum_{z\text{-links}} \sigma_i^z \sigma_j^z. \tag{2}$$

We can also add a time-reversal and parity-breaking potential to this Hamiltonian which comes from third order perturbation theory of a weak magnetic field. The effective potential is $V = \kappa \sum_p V_p$ where $\kappa$ is a coupling constant and the sum is over the plaquettes of the system with each hexagonal plaquette making the following contribution to the potential:

$$V_p = \sigma_6^y \sigma_1^z \sigma_2^x + \sigma_1^x \sigma_2^y \sigma_3^z + \sigma_2^z \sigma_3^x \sigma_4^y + \sigma_3^y \sigma_4^z \sigma_5^x + \sigma_4^x \sigma_5^y \sigma_6^z + \sigma_5^z \sigma_6^y \sigma_1^x, \tag{3}$$

where the sites of the plaquette $p$ have been numbered as in figure 1. Hence, the full Hamiltonian of the model is

$$H = H_0 + \kappa \sum_p V_p. \tag{4}$$

We can define a vortex operator $W_p$ for each plaquette $p$ of the lattice. If we number the sites of the plaquette $p$ as in figure 1, then $W_p$ is defined as

$$W_p = \sigma_1^z \sigma_2^y \sigma_3^x \sigma_4^z \sigma_5^y \sigma_6^x. \tag{5}$$

The vortex operators $W_p$ commute mutually and with the full Hamiltonian, including the time-reversal and parity-breaking potential terms. Consequently the Hilbert space can be written as $\mathcal{H} = \bigoplus_{\{w_p\}} \mathcal{H}_{\{w_p\}}$, where $\mathcal{H}_{\{w_p\}}$ is the common eigenspace of the $W_p$ operators corresponding to the particular configuration of eigenvalues $\{w_p\}$ where $w_p = \pm 1$. We say that a vortex occupies the plaquette $p$ if $w_p = -1$ [1, 9].

Kitaev [1] solved the system by a reduction to free fermions in a static $\mathbb{Z}_2$ gauge field. He showed that the model exhibits four distinct topological phases including three Abelian toric code phases $A_x$, $A_y$, $A_z$, satisfying one and only one of the inequalities:

$$\begin{aligned} A_x &: |J_y| + |J_z| \leqslant |J_x|, \\ A_y &: |J_x| + |J_z| \leqslant |J_y|, \\ A_z &: |J_x| + |J_y| \leqslant |J_z|, \end{aligned} \tag{6}$$

and an additional phase $B$ which occurs when all three inequalities above are not satisfied simultaneously. In the absence of the magnetic field the $B$ phase is gapless, but in the presence of a magnetic field it acquires a gap and becomes the non-Abelian Ising phase. Its quasi-particle excitations, which in our representation are formed by Majorana fermions attached to vortices, show non-Abelian fractional statistics and are known as Ising anyons.

As described in [6], the model can be mapped onto a square lattice whose vertices carry effective spins and hardcore bosons. In this representation, the Hamiltonian of the model (2) acquires the following form

$$\begin{aligned} H_0 = &-J_x \sum_q (b_q^\dagger + b_q)\tau_{q+n_x}^x (b_{q+n_x}^\dagger + b_{q+nx}) \\ &- J_y \sum_q \mathrm{i}\tau_q^z (b_q^\dagger - b_q)\tau_{q+n_y}^y (b_{q+n_y}^\dagger + b_{q+ny}) \\ &- J_z \sum_q (I - 2b_q^\dagger b_q), \end{aligned} \tag{7}$$

where the $\tau_q^\alpha$ are the Pauli operators for the effective spin at a site $q$ and $b_q^\dagger$ and $b_q$ are creation and annihilation operators for hardcore bosons. The sums in the Hamiltonian are over all the sites of the lattice. The contribution to the potential $V$ of a plaquette $P$ has the following form in this representation

$$\begin{aligned} V_P = \tau_d^y\tau_a^z\tau_b^x(I - 2N_a)(b_d^\dagger + b_d)(b_b^\dagger + b_b) &+ \mathrm{i}\tau_b^x(b_a^\dagger + b_a)(b_b^\dagger - b_b) \\ +\tau_b^z\tau_c^y(b_b^\dagger + b_b)(b_c^\dagger + b_c) \qquad &+ \mathrm{i}\tau_b^z\tau_c^z(b_b^\dagger - b_b)(b_d^\dagger + b_d) \\ +\mathrm{i}\tau_c^x(b_c^\dagger + b_c)(b_d^\dagger - b_d) \qquad &+ \tau_d^y\tau_a^z(b_d^\dagger - b_d)(b_a^\dagger - b_a). \end{aligned} \tag{8}$$

The vortex operator $W_P$ for each plaquette $P$ of the lattice is now defined as

$$W_P = (I - 2N_q)(I - 2N_{q+n_y})\tau_q^z\tau_{q+n_y}^y\tau_{q+n_x+n_y}^z\tau_{q+n_x}^y, \tag{9}$$

where $N_q = b_q^\dagger b_q$ is the boson number operator. We say $P$ is occupied by a vortex if the eigenvalue of the corresponding vortex operator is $-1$ and is empty otherwise.

## 3. Lattices on higher genus surfaces

We will now discuss the construction of lattices on closed surfaces with different topologies. To define the model on a closed surface of a higher genus, $g > 1$, we necessarily have to consider a lattice with an Euler characteristic $\chi$ that is negative due to the relation $\chi = 2 - 2g$. A perfect square lattice with the number of vertices $V = N$, the number of plaquettes or faces $F = N$ and the number of edges $E = 2N$ permits at most a closed surface of genus $g = 1$ as its Euler characteristic $\chi = V - \mathrm{E} + F$ is zero. To construct a lattice with negative Euler characteristic from a square lattice requires alterations of some of its vertices or plaquettes. For example, they may include increasing or decreasing the number of edges connected to some vertices or changing the number of edges associated with some plaquettes. We refer to such alterations as defects and emphasize that these are local lattice defects as opposed to non-local defects such as lines of dislocations [10, 11, 12]. We can think of these defects as particles, called genons [13, 14].

We first construct a lattice with $g = 2$ before considering lattices of higher genus. We start with an octagonal piece of square lattice and identify or glue its opposing boundaries together in a way similar to creating a torus by identifying opposite boundaries of a rectangle. The construction is illustrated in figure 2. If we tessellate an octagon with a square lattice and identify the sites residing on the boundary as indicated in figure 3, the resultant lattice will be of genus $g = 2$. We now have a defect plaquette with 12 edges centered around the corners of the original octagon, which are all identified once the boundary edges are glued together. Clearly we could tessellate an octagon with a variety of square lattices of different sizes. The particular lattice we use is characterized by three numbers $\{N_a, N_b, N_c\}$ which specify the number of vertices along the vertical, diagonal and horizontal edges respectively as shown in figure 3. The total number of vertices on such a lattice with dimensions $\{N_a, N_b, N_c\}$ is $N_{\text{tot}} = 2N_b(N_a + N_b + N_c) + N_a N_c$. We can calculate the Euler characteristic by noticing that there are exactly $2N_{\text{tot}}$ edges and $N_{\text{tot}} - 2$ plaquettes including the defect plaquette. Hence we have $\chi = N_{\text{tot}} - 2N_{\text{tot}} + N_{\text{tot}} - 2 = -2$ as desired. We note for completeness that there are other ways of gluing the edges of an octagon together in order to produce a $g = 2$ surface but these may lead to the emergence of undesired line defects. Our approach described above avoids this issue.

Alternatively, we could consider a similar construction using the dual lattice. While the original lattice has each vertex four-valent and all plaquettes are square except the defect plaquette, the dual lattice has all plaquettes square and all vertices four-valent except one defect vertex which is twelve-valent. However, this would require changes of the Hamiltonian of the model. We therefore prefer to work with the original lattice which preserves the form of the Hamiltonian. We will, nevertheless, need to define the vortex operator for the defect plaquette and its magnetic contribution.

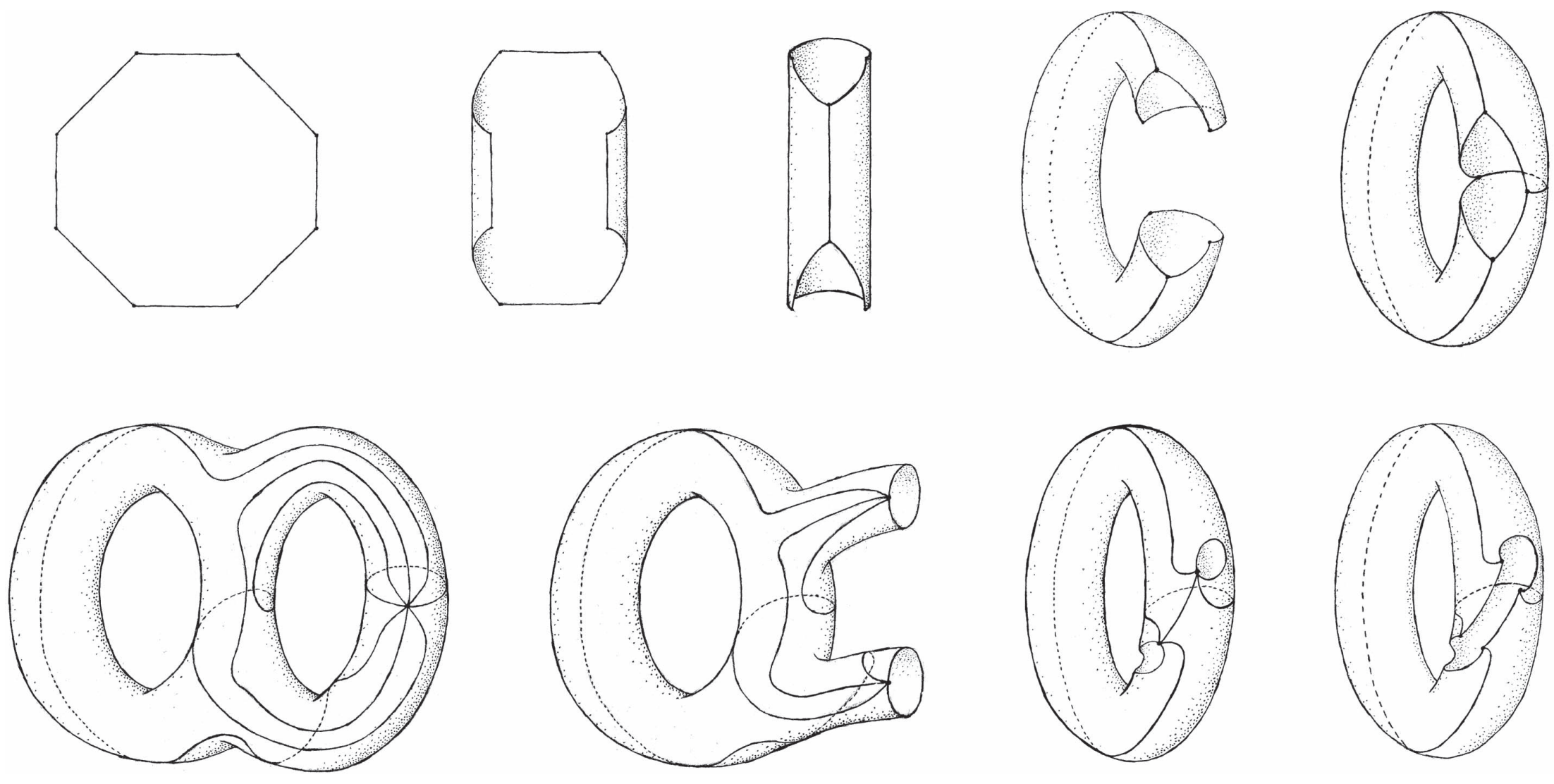

**Figure 2.** To construct a genus 2 surface we can identify the diametrically opposed edges of an octagon.

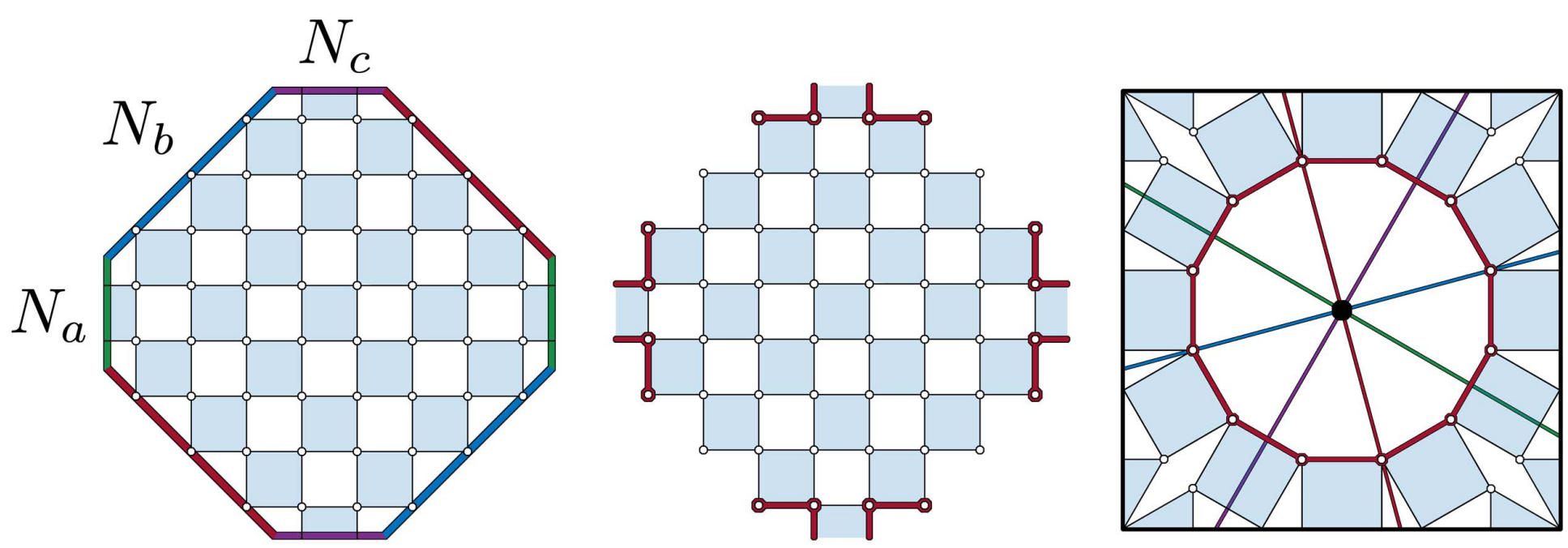

**Figure 3.** The lattices we will be considering on genus 2 surfaces will tessellate an octagon as depicted on the left. They are characterized by three numbers $N_a$, $N_b$ and $N_c$. $N_a$ is the number of links crossing the vertical (green) edge of the octagon. $N_b$ is the number of sites living on a diagonal (blue or red) edge. $N_c$ is the number of links crossing the horizontal (purple) edge of the octagon. When the edges have been identified appropriately, the links colored red in the center image form a closed chain depicted in the image on the right. The corners of the octagon all meet at a common point represented by the black dot at the center of the right image. The corresponding plaquette, centered around this point, will have 12 edges and we will refer to it as the defect plaquette.

We now consider the construction of lattices on surfaces with genus $g > 2$. One approach to generalize the construction developed for the $g = 2$ surfaces above would be to start with a polygon with a greater number of sides (e.g. dodecagon for a $g = 3$ surface) and then glue the opposite sides accordingly. Here we prefer a different and more modular approach which lends itself more naturally to a numerical implementation.

Consider the octagonal piece of lattice as described above. Once all but two of the edges have been glued together, we are left with a lattice with the topology of a torus with two punctures in it. We now use this as a building block for constructing lattices with higher genus. Consider $g - 1$ copies of a torus with two punctures. We can always glue the punctures together in such a way that results in a closed connected surface of genus $g$. With regards to the lattice, we start with $g - 1$ copies of the octagonal piece of square lattice described above and stitch them together to form a chain of octagons as depicted in figure 4. We now form a lattice on a surface of the desired topology by identifying the remaining opposing edges of each octagon as well as by gluing together the remaining edges of the first and the last octagon of the chain. The resultant lattice will have $g - 1$ defect plaquettes, identical to the one described above, located where two octagons are joined together.

We now verify the Euler characteristic for our higher genus lattices. If each octagonal piece has dimensions $N_a$, $N_b$ and $N_c$ then the total number of vertices on this lattice is $(g - 1)N_{\text{tot}}$. We can still uniquely associate every vertex to two edges so the lattice has $2(g - 1)N_{\text{tot}}$ edges. To write down the number of plaquettes as a function of the lattice dimensions $\{N_a, N_b, N_c\}$, we can associate every vertex to the upper right hand plaquette it forms a corner of. Every square plaquette will be assigned a unique vertex while the defect plaquettes will be assigned three. So the number of plaquettes on the lattice is equal to the number of vertices minus 2 for every defect: $(g-1)N_{\text{tot}} - 2(g-1)$. Hence the Euler characteristic of the lattice is $\chi = (g-1)N_{\text{tot}} - 2(g-1)N_{\text{tot}} + (g-1)N_{\text{tot}} - 2(g-1) = 2 - 2g$ as expected.

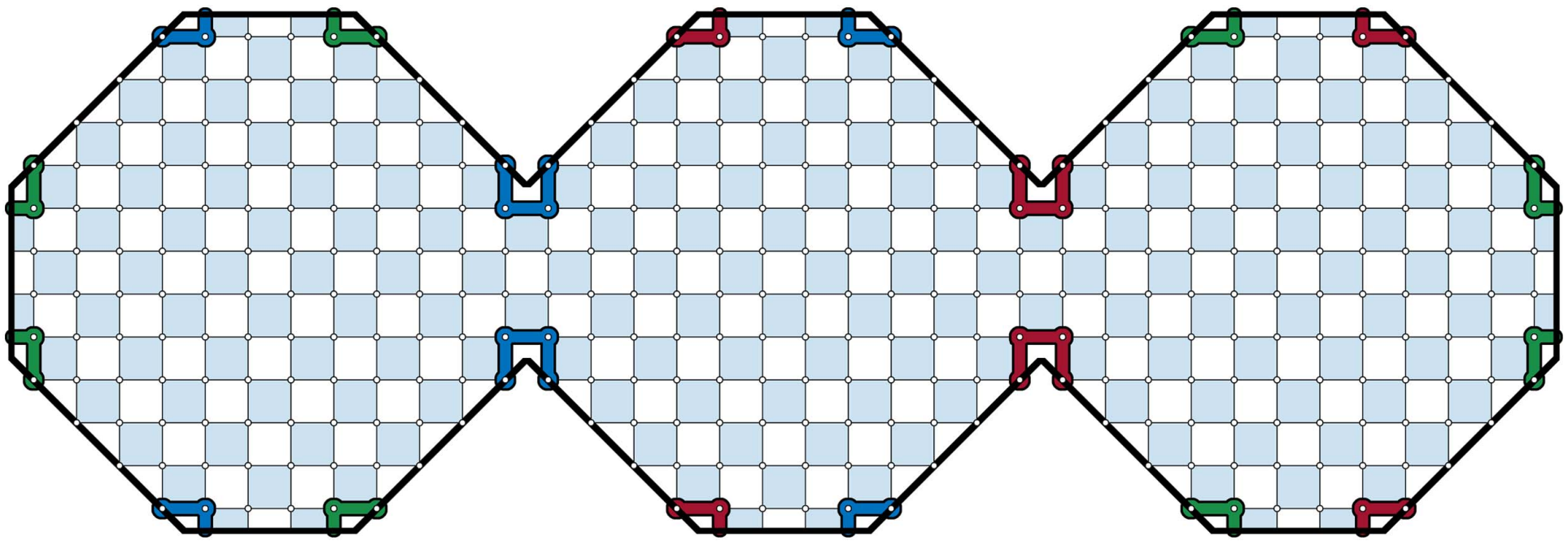

**Figure 4.** Joining three copies of the octagonal piece of lattice as depicted and imposing the same boundary conditions on the diagonal and horizontal edges as described in figure 3 on each octagon results in a lattice tiling a surface of genus $g = 4$ after the vertical edges have been identified.

## 4. The model on surfaces of genus $g \geqslant 2$

We now consider the model on the lattices constructed in the last section. We first write down and discuss the Hamiltonian for the system and its symmetries in the effective spin/hardcore-boson representation of the model. We then fermionize the bosons to obtain a Hamiltonian quadratic in fermionic operators. Since the lattices we will be considering do not have any translational symmetries, we will not be able to write down the ground state in closed form as was done in [6] for the model on a torus. However, the formalism allows one to efficiently diagonalise the Hamiltonian numerically within any particular common eigen-subspace of the models symmetries.

The Hamiltonian in the effective spin/hardcore-boson representation on the lattice described above is of the same form as that on a lattice without defects (2). In both cases, every vertex is four-valent with two horizontal ($x$-links) and two vertical ($y$-links) edges attached. If we denote a site of the lattice by $q$, then by $q + n_x$ we denote the neighbor to the right of $q$ that is connected to it by an $x$-link. Similarly, we use the notation $q + n_y$ to denote the neighbor above $q$ that is connected to it by a $y$-link. The bare Hamiltonian can then be written as follows:

$$\begin{aligned} H_0 = &-J_x \sum_q (b_q^\dagger + b_q)\tau^x_{q+n_x}(b^\dagger_{q+n_x} + b_{q+nx}) \\ &- J_y \sum_q \mathrm{i}\tau^z_q(b_q^\dagger - b_q)\tau^y_{q+n_y}(b^\dagger_{q+n_y} + b_{q+ny}) \\ &- J_z \sum_q (I - 2b_q^\dagger b_q). \end{aligned} \tag{10}$$

Regarding the potential $V = \kappa \sum_P V_P$, the contribution from the square plaquettes are still given by the expression (8). On the other hand, the contribution of the defect plaquettes to the potential are more complicated and are given by

$$\begin{aligned} V_P = &\tau^y_l\tau^z_a\tau^x_b(I - 2N_a)(b_l^\dagger + b_l)(b_b^\dagger + b_b) &&+\mathrm{i}\tau^x_b(b_a^\dagger + b_a)(b_b^\dagger - b_b) \\ &+\tau^z_b\tau^y_c(b_b^\dagger + b_b)(b_c^\dagger + b_c) &&+\mathrm{i}\tau^z_b\tau^z_c(b_b^\dagger - b_b)(b_d^\dagger + b_d) \\ &+\mathrm{i}\tau^x_c(b_c^\dagger + b_c)(b_d^\dagger - b_d) &&+\tau^y_d\tau^z_e(b_d^\dagger - b_d)(b_e^\dagger - b_e) \\ &+\tau^y_d\tau^z_e\tau^x_f(I - 2N_e)(b_d^\dagger + b_d)(b_f^\dagger + b_f) &&+\mathrm{i}\tau^x_f(b_e^\dagger + b_e)(b_f^\dagger - b_f) \\ &+\tau^z_f\tau^y_g(b_f^\dagger + b_f)(b_g^\dagger + b_g) &&+\mathrm{i}\tau^z_f\tau^z_g(b_f^\dagger - b_f)(b_h^\dagger + b_h) \\ &+\mathrm{i}\tau^x_g(b_g^\dagger + b_g)(b_h^\dagger - b_h) &&+\tau^y_h\tau^z_i(b_h^\dagger - b_h)(b_i^\dagger - b_i) \\ &+\tau^y_h\tau^z_i\tau^x_j(I - 2N_i)(b_h^\dagger + b_h)(b_j^\dagger + b_j) &&+\mathrm{i}\tau^x_j(b_i^\dagger + b_i)(b_j^\dagger - b_j) \\ &+\tau^z_j\tau^y_k(b_j^\dagger + b_j)(b_k^\dagger + b_k) &&+\mathrm{i}\tau^z_j\tau^z_k(b_j^\dagger - b_j)(b_l^\dagger + b_l) \\ &+\mathrm{i}\tau^x_k(b_k^\dagger + b_k)(b_l^\dagger - b_l) &&+\tau^y_l\tau^z_a(b_l^\dagger - b_l)(b_a^\dagger - b_a). \end{aligned} \tag{11}$$

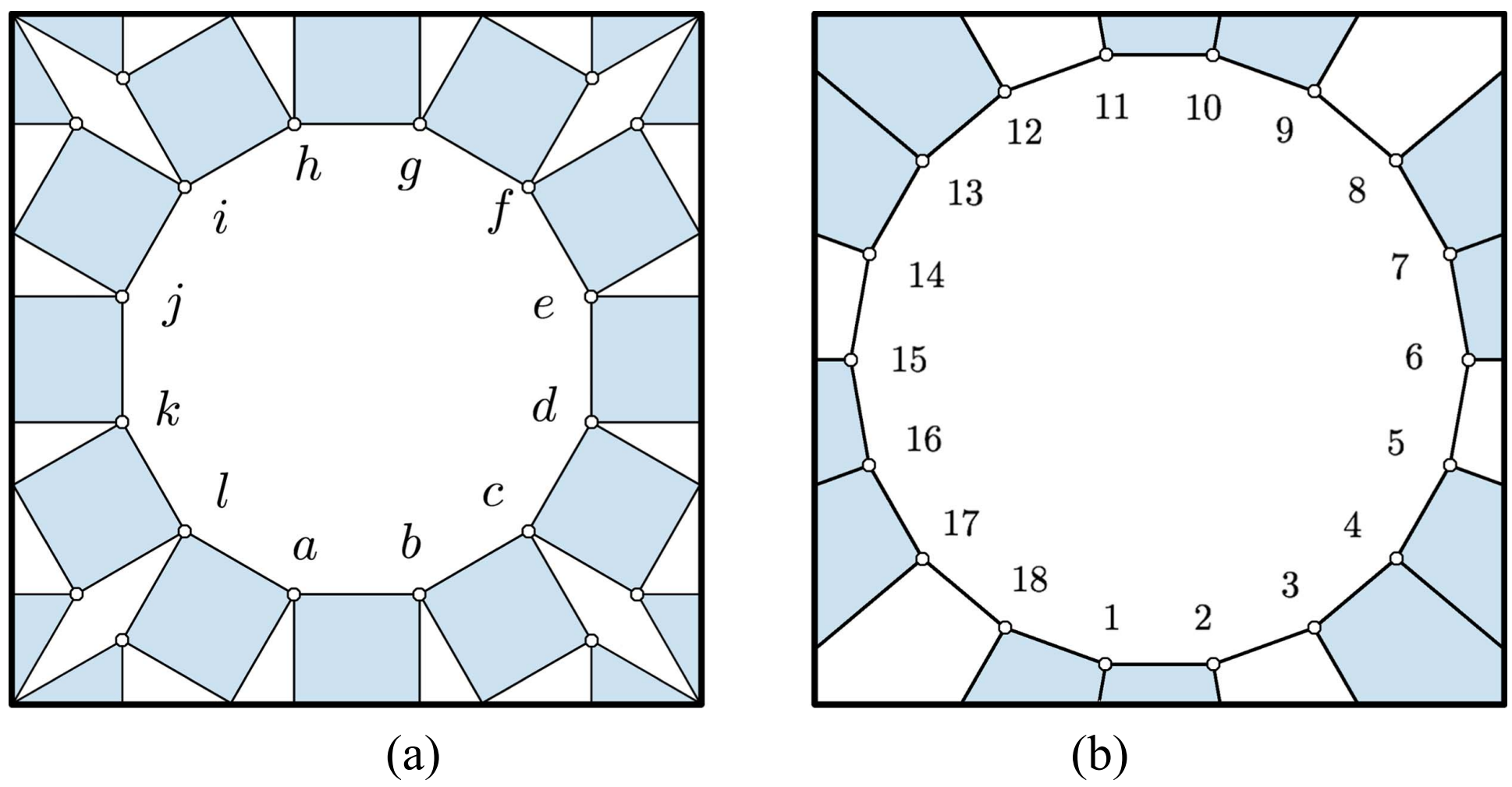

**Figure 5.** The 12 sided defect of the square lattice (left) becomes an eighteen sided defect plaquette (right) in the honeycomb lattice picture.

This expression follows from translating the three-body spin terms, linked at the third order of perturbation theory to the weak magnetic field, into the effective spin/hardcore-boson representation. In the original honeycomb picture, the defect corresponds to a plaquette with eighteen edges and the contribution to the potential by a defect plaquette is the following sum of three-body spin terms:

$$\begin{aligned} V_p = {} & \sigma_{18}^y\sigma_1^z\sigma_2^x + \sigma_1^x\sigma_2^y\sigma_3^z + \sigma_2^z\sigma_3^x\sigma_4^y + \sigma_3^y\sigma_4^z\sigma_5^x + \sigma_4^x\sigma_5^y\sigma_6^z + \sigma_5^z\sigma_6^x\sigma_7^y \\ & + \sigma_6^y\sigma_7^z\sigma_8^x + \sigma_7^x\sigma_8^y\sigma_9^z + \sigma_8^z\sigma_9^x\sigma_{10}^y + \sigma_9^y\sigma_{10}^z\sigma_{11}^x + \sigma_{10}^x\sigma_{11}^y\sigma_{12}^z + \sigma_{11}^z\sigma_{12}^x\sigma_{13}^y \\ & + \sigma_{12}^y\sigma_{13}^z\sigma_{14}^x + \sigma_{13}^x\sigma_{14}^y\sigma_{15}^z + \sigma_{14}^z\sigma_{15}^x\sigma_{16}^y + \sigma_{15}^y\sigma_{16}^z\sigma_{17}^x + \sigma_{16}^x\sigma_{17}^y\sigma_{18}^z + \sigma_{17}^z\sigma_{18}^x\sigma_1^y, \end{aligned} \tag{12}$$

where the sites of the plaquette are numbered as depicted in figure 5.

We can still define a vortex operator which commutes with the full Hamiltonian $H = H_0 + \kappa\sum_P V_P$ for every plaquette. For square plaquettes the vortex operator is defined as in equation (9). For the defect plaquettes however we define the vortex operator as follows

$$W_P = (1 - 2N_a)(1 - 2N_d)(1 - 2N_e)(1 - 2N_h)(1 - 2N_i)(1 - 2N_l)\tau_a^z\tau_b^y\tau_c^z\tau_d^y\tau_e^z\tau_f^y\tau_g^z\tau_h^y\tau_i^z\tau_j^y\tau_k^z\tau_l^y. \tag{13}$$

This definition of the vortex operator for the defect plaquette is equivalent to the following product of Pauli operators in the original honeycomb picture of the model:

$$W_P = \sigma_1^z\sigma_2^y\sigma_3^x\sigma_4^z\sigma_5^y\sigma_6^x\sigma_7^z\sigma_8^y\sigma_9^x\sigma_{10}^z\sigma_{11}^y\sigma_{12}^x\sigma_{13}^z\sigma_{14}^y\sigma_{15}^x\sigma_{16}^z\sigma_{17}^y\sigma_{18}^x. \tag{14}$$

In addition to the vortex operators, we also define an operator, which commutes with the Hamiltonian, for every generator in a basis for the 1st $\mathbb{Z}_2$-homology group $H_1$ of the lattice. We will call these operators loop operators and to define them we will need to choose a basis for $H_1$ and a particular representative from each homology class in that basis. For a lattice of genus $g \geqslant 2$ the rank of $H_1$ is $2g$ so we will need to choose $2g$ homologically distinct cycles. We will choose the cycles depicted in figure 6 and their associated homology classes as the representatives and basis respectively. As depicted, for a lattice with $g - 1$ copies of an octagonal piece of lattice we will choose three cycles on the first copy, two cycles on every other copy (reflecting the fact that every additional copy increases the genus by 1 and the rank of $H_1$ by 2) and one horizontal cycle that spans each octagon. The loop operators we will define for these cycles will act on the sites of the lattice that are connected to the links that constitute the cycles. How a loop operator acts on a particular site is determined by the way the associated cycle passes through it. There are six ways a cycle can pass through a site as depicted in figure 7 and we will associate a single site operator with each of them as follows:

$$\begin{aligned} &\text{Horizontal:} && -(1 - 2N)\tau^x \\ &\text{Vertical:} && \tau^x \\ &\text{Corner 1:} && -(1 - 2N)\tau^y \\ &\text{Corner 2:} && \mathrm{i}\tau^z \\ &\text{Corner 3:} && -\mathrm{i}(1 - 2N)\tau^z \\ &\text{Corner 4:} && -\tau^y. \end{aligned} \tag{15}$$

We define the loop operator for a particular cycle as the composition of all the single site operators associated with the sites it passes through times a minus sign. For example, the cycle numbered 3 has a horizontal part, a

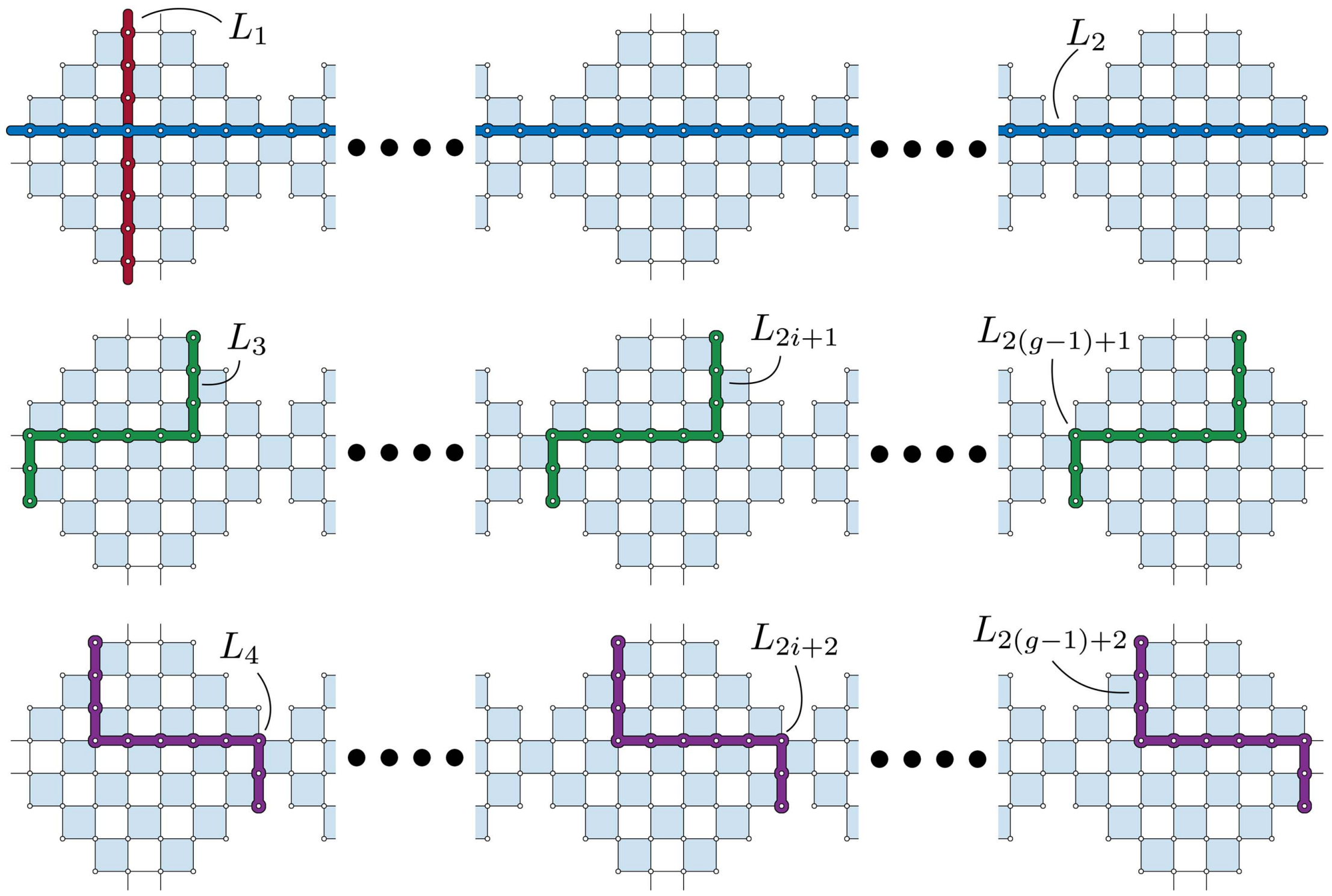

**Figure 6.** The first two cycles we choose, denoted $L_1$ and $L_2$ respectively, wrap vertically around the first octagonal part of the lattice and horizontally around the entire lattice as depicted. Also, for each octagonal part $i = 1 \cdots (g - 1)$, we choose the two cycles labeled by $L_{2i+1}$ and $L_{2i+2}$. There are $2g$ cycles in total.

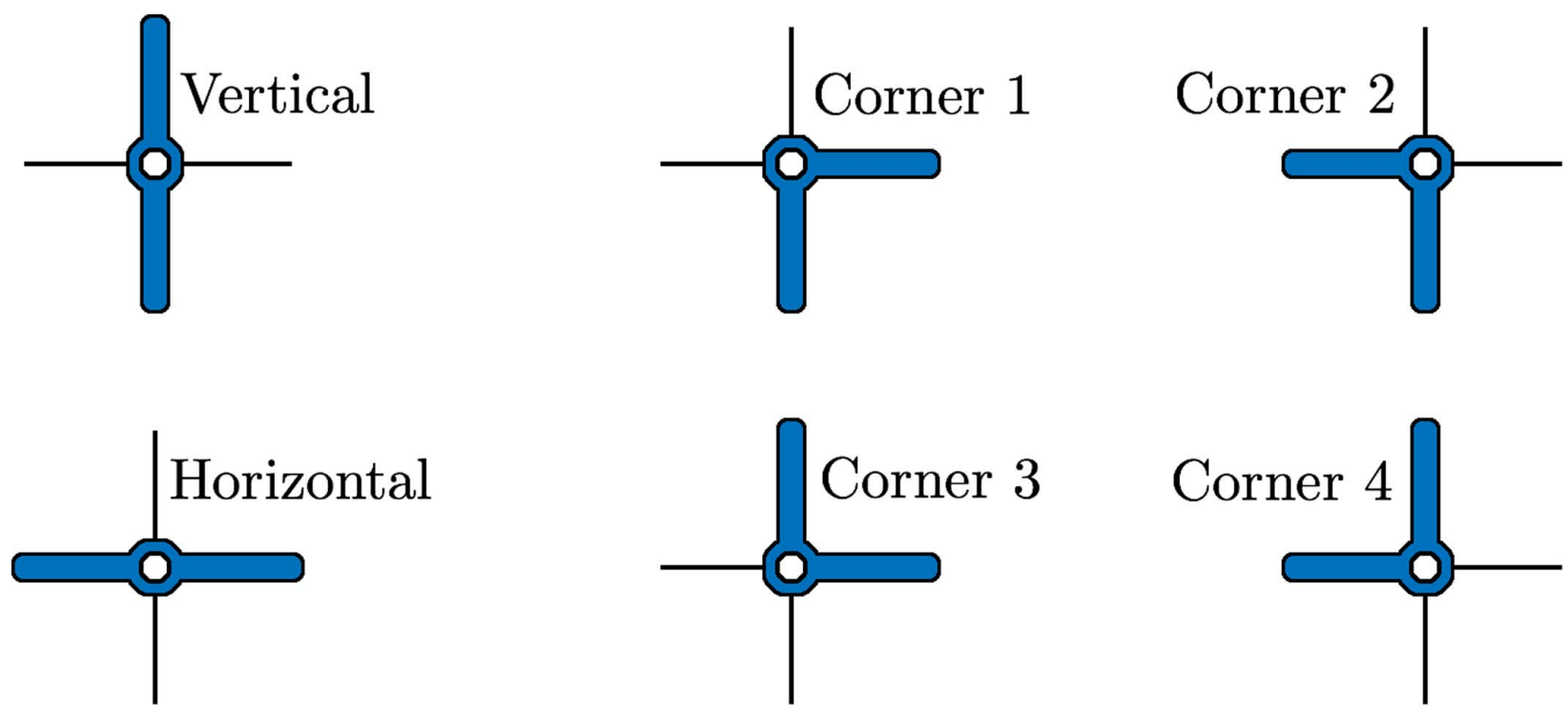

**Figure 7.** If a 1-chain passes through a site once, it can only do so in one of six ways as depicted.

vertical part and two corners and so the loop operator for this cycle can be written as follows:

$$L_3 = -\overbrace{[-(1-2N)\tau^y]}^{\text{Corner 1}}\overbrace{\left[\prod_i -(1-2N_i)\tau_i^x\right]}^{\text{Horizontal}}\overbrace{[-\tau^y]}^{\text{Corner 4}}\overbrace{\left[\prod_i \tau_i^x\right]}^{\text{Vertical}}. \tag{16}$$

One can in principle define a different set of loop operators that commute with the Hamiltonian but these will in general be equivalent to a product of the loop operators already defined times a product of vortex operators [9].

The vortex and loop operators form a set of commuting observables, allowing us to decompose the Hilbert space as follows:

$$\mathcal{H} = \bigoplus_{\{w_p, l_i\}} \mathcal{H}_{\{w_p, l_i\}}. \tag{17}$$

Here $\{w_p\}$ and $\{l_i\}$ denote particular configurations of eigenvalues of all the vortex and loop operators respectively. $\mathcal{H}_{\{w_p, l_i\}}$ is the common eigen-subspace of all vortex and loop operators corresponding to the configuration $\{w_p, l_i\}$. The method we use to solve the model involves restricting the Hamiltonian to one of these subspaces where it can be expressed as a combination of terms that are quadratic in fermionic operators. The restricted Hamiltonian can then be diagonalized by an appropriate Bogoliubov transformation.

We now change to a basis of the Hilbert space which reflects the decomposition (17). It seems natural to consider the common eigenvectors of the vortex and loop operators along with the eigenstates of the boson number operator ($N_q = b_q^\dagger b_q$) for each site $q$ of the square lattice. However, the basis so defined would be overcomplete. If there are $(g-1)N_{\text{tot}}$ sites in the lattice, the model clearly has $2^{2(g-1)N_{\text{tot}}}$ configurations of effective spins and bosons. Yet there are $(g-1)N_{\text{tot}} - 2(g-1)$ vortex operators, $(g-1)N_{\text{tot}}$ boson number operators and $2g$ loop operators, all of which have eigenvalues $\pm 1$. So there are $2^{(g-1)N_{\text{tot}}-2(g-1)} \times 2^{(g-1)N_{\text{tot}}} \times 2^{2g} = 2^{2(g-1)N_{\text{tot}}+2}$ distinct combinations of eigenvalues a common eigenvector of this set of observables might have. However, the vortex and number operators are not completely independent operators as they satisfy two conditions.

The first condition is the fact that the product of all vortex operators is equivalent to the identity operator. That is,

$$\prod_P W_P = 1, \tag{18}$$

where the product is over all the plaquettes of the lattice. Since a product of vortex operators can be thought of as counting the parity of vortices occupying the associated plaquettes, this essentially means there can only be an even number of vortices in total in the model. So the number of independent vortex operators is $(g-1)N_{\text{tot}} - 2(g-1) - 1$ and hence the number of configurations of vortices in the model is $2^{(g-1)N_{\text{tot}}-2(g-1)-1}$.

The second condition is a relation between the parity of bosons in the system and a certain product of vortex operators. For a lattice where the numbers $N_a$ and $N_b$ are both even, we can consider a set of plaquettes forming a checker board pattern as depicted in the top left image of figure 8 by the colored squares. It is easy to check that since the Pauli operators square to the identity, the product of the vortex operators associated with the colored (or uncolored) plaquettes is equivalent to the boson parity operator

$$\prod_{\text{colored}} W_P = \prod_q (1 - 2N_q), \tag{19}$$

where $q$ runs over all the sites of the lattice. In other words, the parity of the number of bosons must be the same as the parity of the number of vortices on colored plaquettes (or equivalently uncolored plaquettes). Since the parity of bosons is fixed to be 1 or $-1$ depending on the configuration of vortices, the number of independent boson number operators $N_q$ is $(g-1)N_{\text{tot}} - 1$ and hence the number of configurations of bosons in the model is $2^{(g-1)N_{\text{tot}}-1}$.

For lattices where $N_a$ or $N_b$ are odd numbers, there is a similar dependence of the boson parity on the configuration of vortices in the system. For such lattices, we cannot color the plaquettes with a perfect checker board pattern but we can consider sets of plaquettes as depicted in figure 8, such that the checker board pattern is misaligned along a 1-cycle of links that separate plaquettes of the same color. The exact pattern we choose for coloring in plaquettes and the associated cycle along which the checker board pattern is misaligned depends on the parity of the numbers $N_a$, $N_b$ and $g$ for the lattice and is described in figure 8. If we compose the corresponding vortex operators, the Pauli operators for sites away from this cycle will cancel out as they did before but along the cycle, the resultant operator will act with a string of Pauli operators and may not act with the parity operator $(1 - 2N_q)$ for some sites. However, we can cancel the action of these Pauli operators, and replace any missing single site parity operators we need to obtain the full boson parity operator, by composing this product of vortex operators with a product of loop operators that act on the sites connected to the links of the cycle. The desired product of loop operators that act on the sites connected to the links of the cycle are shown in figure 8. In general, the boson parity operator can be written as

$$\prod_q (1 - 2N_q) = (-1)^{N_a N_c (g-1)} L_1^{N_a} (L_2 L_3 L_4)^{N_c (g-1)} \left( \prod_{i=1}^{g-1} L_{2i+1} L_{2i+2} \right)^{N_c} \prod_{\text{colored}} W_p, \tag{20}$$

where the product of vortex operators is over the corresponding set of colored plaquettes. These conditions mean we can form a complete set of commuting observables reflecting the decomposition (17) by taking all vortex and loop operators with every single site boson number operator and then excluding one vortex operator and one number operator.

The next step of the solution is to use a 'Jordan–Wigner' type transformation to fermionize the bosons of the model. This should result in a Hamiltonian which is quadratic in fermionic operators which we will then be able to solve using the Bogoliubov-de Gennes (BdG) technique. To fermionize the bosons, we will define a Jordan–Wigner type string operator $S_q$ for each site $q$ of the lattice. The composition of these string operators with the boson creation and annihilation operators will be fermionic creation and annihilation operators. Expressing the Hamiltonian and other observables in terms of these new operators will effectively transform the hardcore bosons of the model into fermions.

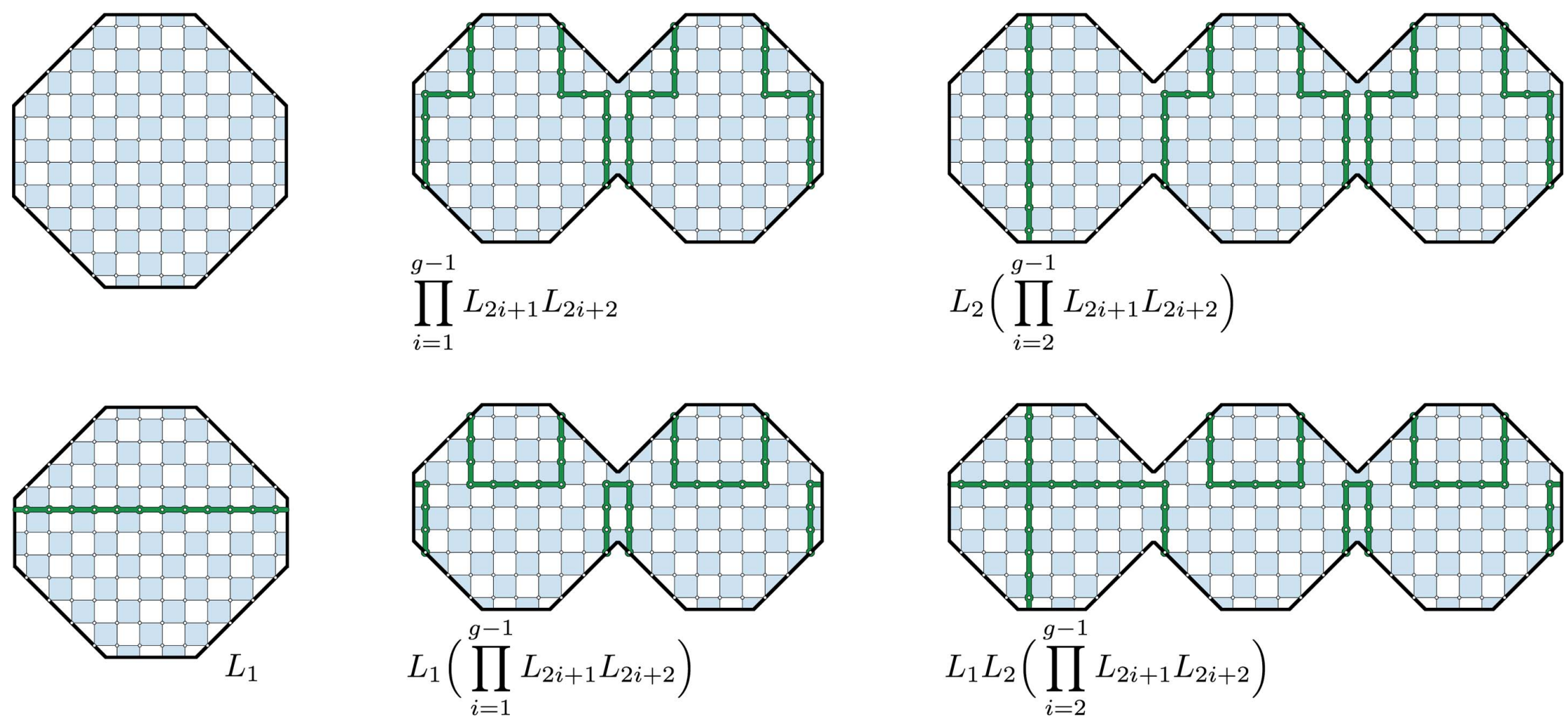

**Figure 8.** If both $N_a$ and $N_b$ are even numbers, then the plaquettes of each octagonal part of the lattice can be colored in a checker board pattern as depicted in the top left image. If $N_a$ is odd but $N_b$ is even we can color the plaquettes of each octagonal part as depicted in the bottom left image. In this case the checker board pattern is misaligned along the cycle associated with the loop operator $L_1$. If $N_b$ is odd, then the pattern we color the plaquettes in depends on the parity of $g$. If $g$ is odd then the plaquettes of each octagonal part can be colored like one of the octagonal parts of the top center image if $N_a$ is even or the bottom center image if $N_a$ is odd. If both $g$ and $N_a$ are even then the plaquettes of the first octagonal part can be colored as depicted in the top right image while those of the other octagonal parts are colored like the other octagonal parts of the top right image. If $g$ is even and $N_a$ is odd then the plaquettes of the first octagonal part can be colored as depicted in the bottom right image while those of the other octagonal parts are colored like the other octagonal parts of the bottom right image.

To define a string operator for a site $q$ of the lattice we consider the following: if we had a particle located at the reference site (as in figure 9(a)) we can always move that particle to any site $q$ by first moving it to the right an appropriate number of sites and then up an appropriate number of sites. Even the sites below the level of the reference site can be reached in this way by making use of the boundary conditions as shown in figure 9(b). We can associate a single site operator for every site traversed in the path just described connecting the reference site to the site $q$. The string operator for $q$, denoted $S_q$, will be defined as the composition of these operators. To every site $i$ crossed by the horizontal part of the path we associate the operator $-(1 - N_i)\tau_i^x$, to the corner of the path we associate the operator $-\tau_i^y$, to every site $i$ crossed by the vertical part of the path we associate the operator $\tau_i^x$ and to the last site of the path we associate the operator $\tau_i^y$. Since each of these operators act on different sites, they all commute with each other and so we are free to define $S_q$ as the composition of these operators without worrying about the order of composition. If we let $q_x$ denote the number of sites that need to be traversed in the horizontal part of the path with the site at the corner and $q_y$ the number of sites that need to be traversed in the vertical part of the path with the site at the end, then we can number the sites of the path from 1 to $q_x + q_y$, beginning at the reference site and ending at $q$ and we can write the string operator for $q$ as follows:

$$\begin{aligned} S_q = [-(1 - N_1)\tau_1^x] \times \cdots \times [-(1 - N_{q_x-1})\tau_{q_x-1}^x] \quad & \text{horizontal part} \\ \times -\tau_{q_x}^y \quad & \text{corner} \\ \times \tau_{q_x+1}^x \times \cdots \times \tau_{q_x+q_y-1}^x \quad & \text{vertical part} \\ \times \tau_{q_x+q_y}^y. \quad & \text{end} \end{aligned} \tag{21}$$

If we consider two string operators $S_q$ and $S_{q'}$ such that $q \neq q'$ there will be a single site, shared by the paths defining the string operators, where the action of $S_q$ anti-commutes with the action of $S_{q'}$. It follows that composing the string operator $S_q$ with the bosonic creation and annihilation operators for the site $q$ defines fermionic creation and annihilation operators for $q$ which we denote by $c_q^\dagger$ and $c_q$.

$$c_q^\dagger \equiv b_q^\dagger S_q,\ c_q \equiv b_q S_q, \tag{22}$$

$$\{c_q^\dagger, c_{q'}\} = \delta_{q,q'},\ \{c_q^\dagger, c_{q'}^\dagger\} = 0,\ \{c_q, c_{q'}\} = 0. \tag{23}$$

Expressing the basic Hamiltonian in terms of these fermionic creation and annihilation operators yields the following sum of quadratic fermionic terms,

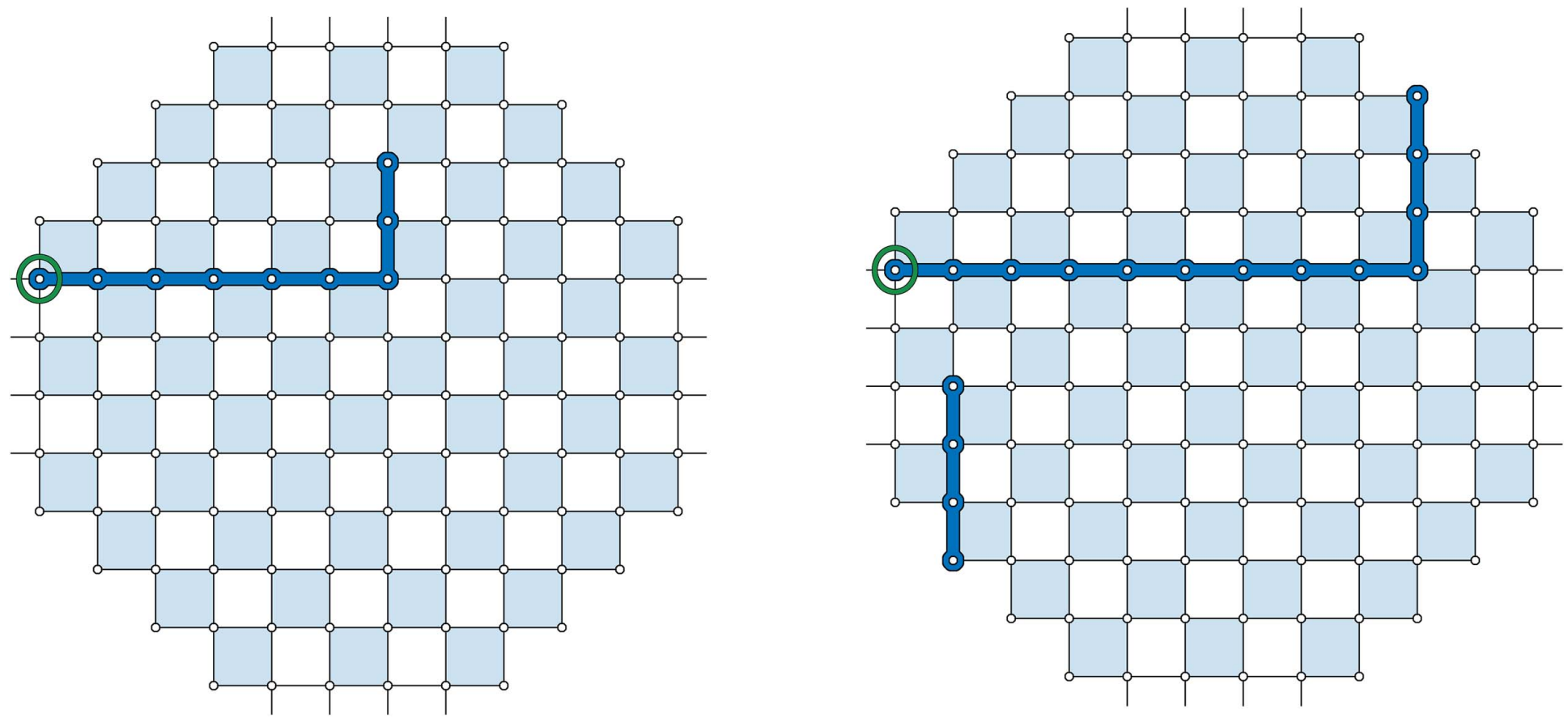

**Figure 9.** The reference site we have chosen in defining string operators for each site of the lattice is encircled in green. Any site can be reached from the reference site by moving to the right a number of sites and then moving up a number of sites as shown in (a). Sites beneath the reference site can be reached in this way if the boundary conditions of the lattice are utilized as shown in (b).

$$\begin{aligned} H_0 = J_x \sum_{x\text{-links}} X_{q1,q2}(c_{q1}^\dagger - c_{q1})(c_{q2}^\dagger + c_{q2}) \\ + J_y \sum_{y\text{-links}} Y_{q1,q2}(c_{q1}^\dagger - c_{q1})(c_{q2}^\dagger + c_{q2}) \\ + J_z \sum_q (2N_q - I), \end{aligned} \tag{24}$$

where, if $q1$ and $q2$ are sites on the left and right hand side of a $x$-link respectively, $X_{q1,q2} = -(I - 2N_{q1})S_{q1}\tau_{q2}^x S_{q2}$ and, if $q1$ and $q2$ are sites at the bottom and top of a $y$-link respectively, $Y_{q1,q2} = \mathrm{i}\tau_{q1}^z S_{q1}\tau_{q2}^y S_{q2}$. Note, the string operators square to the identity and so the number operator $N_q = b_q^\dagger b_q = c_q^\dagger c_q$ is unchanged when expressed in fermion operators and hence the vortex operators are also left unchanged.

Noting that both the $X_{q,q'}$ and $Y_{q,q'}$ operators, being products of string operators, act on a closed loop of sites, we can associate a 1-cycle with each of the $X_{q,q'}$ and $Y_{q,q'}$ operators, namely the set of links joining the sites being acted on. These operators will always be equivalent to a product of loop operators, which is determined by the homology class of this cycle, and a product of vortex operators which is determined by a certain 2-chain related to the homology class of the cycle and the representatives of the homology classes we have chosen as a basis for $H_1$. Recall that each loop operator is associated to a non-trivial cycle, the collection of which represent the generators of $H_1$. So whatever the homology class may be for the cycle $a$ associated with an $X$ or $Y$ operator, we can always create a unique cycle $b$ which will be homologous to $a$ by adding some combination of the cycles associated with the loop operators. A particular $X$ or $Y$ operator is proportional to the product of the loop operators corresponding to the cycles used in the combination forming $b$.

There will also be a 2-chain, which we denote by $\varsigma$, which will have $a + b$ as a boundary. A particular $X$ or $Y$ operator is also proportional to a product of the vortex operators associated with the plaquettes which constitute $\varsigma$. We note that while such a 2-chain $\varsigma$ is not unique, the operator obtained by multiplying the vortex operators associated with the plaquettes of the 2-chain $\varsigma$ is unique. For example, if we cut out a cylinder with boundaries $a$ and $b$ from a torus, the product of the vortex operators inside of the cylinder is the same as in its complement. This follows from the fact that vortex operators square to the identity and the relation (18). In general, when expressed in terms of loop and vortex operators, the $X$ and $Y$ operators are of the same form. We will use a notation to reflect this by letting $Z_q$ denote $X_q$ if $q$ is a $x$-link and $Y_q$ if $q$ is a $y$-link. Explicitly we have

$$Z_q = L_1^{a_1} \cdots L_{2g}^{a_{2g}} \prod_{p \in \varsigma(q)} W_p, \tag{25}$$

where $(a_1, \cdots, a_{2g}) \in H_1$ is the homology class of the cycle associated with the link $q$ described above. So the Hamiltonian can be written as follows

$$H_0 = J_x \sum_{x\text{-links}} Z_{q1,q2}(c_{q1}^\dagger - c_{q1})(c_{q2}^\dagger + c_{q2}) + J_y \sum_{y\text{-links}} Z_{q1,q2}(c_{q1}^\dagger - c_{q1})(c_{q2}^\dagger + c_{q2}) + J_z \sum_q (2N_q - I). \tag{26}$$

Since the basic Hamiltonian is quadratic in fermionic operators, it can be written using the BdG formalism:

$$H = \frac{1}{2}[c^\dagger \quad c]\begin{bmatrix} \xi & \Delta \\ \Delta^\dagger & -\xi^T \end{bmatrix}\begin{bmatrix} c \\ c^\dagger \end{bmatrix}, \tag{27}$$

where the elements of the $(g-1)N_{\text{tot}} \times (g-1)N_{\text{tot}}$ matrices $\xi$ and $\Delta$ are given by

$$\xi_{q,q'} = J_z\delta_{q,q'} + J_x Z_{q,q'}(\delta^{(x)}_{q,q'} + \delta^{(x)}_{q',q}) + J_y Z_{q,q'}(\delta^{(y)}_{q,q'} + \delta^{(y)}_{q',q}),$$
$$\Delta_{q,q'} = J_x Z_{q,q'}(\delta^{(x)}_{q,q'} - \delta^{(x)}_{q',q}) + J_y Z_{q,q'}(\delta^{(y)}_{q,q'} - \delta^{(y)}_{q',q}). \tag{28}$$

Here, $\delta_{q,q'}$ is the usual Kronecker delta and $\delta^{(x)}_{q,q'}$ is defined to be 1 if $q$ and $q'$ are the sites on the left and right hand side of an $x$-link respectively and zero otherwise. Similarly, $\delta^{(y)}_{q,q'}$ is 1 if $q$ and $q'$ are the sites on the bottom and top side of a $y$-link respectively and zero otherwise.

Regarding the potential, when expressed in terms of the fermionic creation and annihilation operators, each term appearing in the sum defining the contribution from a plaquette inherits a product of string operators similar to the $X$ and $Y$ operators. The potential also becomes quadratic in fermionic operators and can be written as

$$V = \kappa \sum_p V_p = \frac{\kappa}{2}[c^\dagger \quad c]\begin{bmatrix} \bar{\xi} & \bar{\Delta} \\ \bar{\Delta}^\dagger & \bar{\xi}^T \end{bmatrix}\begin{bmatrix} c \\ c^\dagger \end{bmatrix}, \tag{29}$$

where the elements of the matrices $\bar{\xi}$ and $\bar{\Delta}$ are given by

$$\bar{\xi}_{q,q'} = \mathrm{i}\sum_\rho Z_{q,\rho}Z_{\rho,q'}(-\delta^{(x)}_{q,\rho}\delta^{(y)}_{q',\rho} + \delta^{(x)}_{q',\rho}\delta^{(y)}_{q,\rho} + \delta^{(x)}_{\rho,q'}\delta^{(y)}_{\rho,q} - \delta^{(x)}_{\rho,q}\delta^{(y)}_{\rho,q'}), \tag{30}$$

and

$$\bar{\Delta}_{q,q'} = \mathrm{i}\sum_\rho Z_{q,\rho}Z_{\rho,q'}(\delta^{(x)}_{q,\rho}\delta^{(y)}_{q',\rho} - \delta^{(x)}_{q',\rho}\delta^{(y)}_{q,\rho} + \delta^{(x)}_{\rho,q'}\delta^{(y)}_{\rho,q} - \delta^{(x)}_{\rho,q}\delta^{(y)}_{\rho,q'}) - 2\mathrm{i}Z_{q,q'}(\delta^{(x)}_{q,q'} - \delta^{(x)}_{q',q}) + 2\mathrm{i}Z_{q,q'}(\delta^{(y)}_{q,q'} - \delta^{(y)}_{q',q}). \tag{31}$$

Hence, the full Hamiltonian is

$$H = \frac{1}{2}[c^\dagger \quad c]\begin{bmatrix} \xi + \kappa\bar{\xi} & \Delta + \kappa\bar{\Delta} \\ \Delta^\dagger + \kappa\bar{\Delta}^\dagger & -\xi^T - \kappa\bar{\xi}^T \end{bmatrix}\begin{bmatrix} c \\ c^\dagger \end{bmatrix}. \tag{32}$$

Now, if we were to choose to restrict or attention a particular common eigenspace of the vortex and loop operators appearing in (17), we may replace the vortex and loop operators appearing in the definition of the $Z$ operators by their eigenvalues. This would result in replacing the $Z$ operators appearing in the Hamiltonian by their eigenvalues and we could diagonalise the square matrix appearing (32) numerically to study the spectrum in that subspace.

## 5. Calculating the ground state

Once we have restricted the Hamiltonian (27) to a particular homology sector with no vortices, diagonalising the BdG matrix results in a unitary matrix $T$ whose columns hold the coefficients of quasi-particle excitations in the $c$-fermion basis. So we have

$$H = \frac{1}{2}[c^\dagger \quad c]\begin{bmatrix} \xi + \kappa\bar{\xi} & \Delta + \kappa\bar{\Delta} \\ \Delta^\dagger + \kappa\bar{\Delta}^\dagger & -\xi^T - \kappa\bar{\xi}^T \end{bmatrix}\begin{bmatrix} c \\ c^\dagger \end{bmatrix} \tag{33}$$

$$= \frac{1}{2}[\gamma^\dagger \quad \gamma]\begin{bmatrix} E & \\ & -E \end{bmatrix}\begin{bmatrix} \gamma \\ \gamma^\dagger \end{bmatrix} \tag{34}$$

$$= \frac{1}{2}\sum_i (E_i\gamma_i^\dagger\gamma_i - E_i\gamma_i\gamma_i^\dagger) \tag{35}$$

$$=\sum_i E_i \gamma_i^\dagger \gamma_i - \frac{1}{2}\sum_i E_i, \tag{36}$$

with

$$\begin{bmatrix} \gamma \\ \gamma^\dagger \end{bmatrix} \equiv [T^\dagger] \begin{bmatrix} c \\ c^\dagger \end{bmatrix}. \tag{37}$$

Here, $E$ is a real $((g-1)N_{\text{tot}}) \times (g-1)(N_{\text{tot}})$ diagonal matrix with non-negative eigenvalues. We note the spectrum is symmetric about zero due to the particle–hole symmetry of BdG Hamiltonians. The ground state of the system in a particular vortex/homology sector is the quasi-particle vacuum defined by the property that it is annihilated by all quasi-particle annihilation operators (which we can number 1 to $(g-1)N_{\text{tot}}$)

$$\gamma_i|\text{GS}\rangle = 0, \quad \text{for all } i = 1, .., (g-1)N_{\text{tot}}. \tag{38}$$

If we denote the $c$-fermion vacuum by $|-\rangle$, it is easy to check that the state $|\phi\rangle = \mathcal{N}\prod_k \gamma_k|-\rangle$ satisfies the above condition, where the product runs over all the occupied quasi-particle modes of $|-\rangle$ and $\mathcal{N}$ is a normalizing constant. This state may or may not be the true ground state of the system depending on the parity of the number of occupied $c$-fermions modes it has. We need to make sure the $c$-fermion parity of $|\phi\rangle$ satisfies the condition (20) for the particular vortex/homology sector we have restricted to. If this condition is satisfied, $|\phi\rangle$ is the true ground state of the system. If the condition is not satisfied, $|\phi\rangle$ represents an unphysical state. However, we can rectify the situation by applying $\gamma_1^\dagger$, the minimum positive energy quasi-particle creation operator, to $|\phi\rangle$ to create the true ground state. Hence, if we denote the ground state by $|\text{GS}\rangle$, we have

$$|\text{GS}\rangle = \begin{cases} |\phi\rangle = \mathcal{N}\prod_k \gamma_k|-\rangle, & \text{if } |\phi\rangle \text{ satisfies (20)}, \\ \gamma_1^\dagger|\phi\rangle, & \text{otherwise.} \end{cases} \tag{39}$$

It follows from (36) and (39) that the ground state energy for whichever homology/vortex sector we have restricted to is

$$E_{\text{GS}} = \begin{cases} -\frac{1}{2}\sum_i E_i, & \text{if } |\phi\rangle \text{ satisfies (20)}, \\ E_1 - \frac{1}{2}\sum_i E_i, & \text{otherwise.} \end{cases} \tag{40}$$

To be able to calculate the ground state energy numerically for a particular vortex/homology sector, we need to understand how the matrix $T$ represents the state $|\phi\rangle$ and how it can tell us the parity of the of the number of occupied $c$-fermions modes it has. In general, $T$ will be of the following form:

$$T = \begin{bmatrix} U & V^\star \\ V & U^\star \end{bmatrix}, \tag{41}$$

where $U$ and $V$ are $(g-1)N_{\text{tot}} \times (g-1)N_{\text{tot}}$ matrices which, since $T$ must be unitary, must satisfy

$$U^\dagger U + V^\dagger V = 1, UU^\dagger + V^\star V^T = 1, \tag{42}$$

$$U^T V + V^T U = 0, UV^\dagger + V^\star U^T = 0. \tag{43}$$

Bloch and Messiah were able to show that a unitary matrix of the form (41) can be decomposed as follows [15, 16]

$$T = \begin{bmatrix} U & V^\star \\ V & U^\star \end{bmatrix} = \begin{bmatrix} D & \\ & D^\star \end{bmatrix} \begin{bmatrix} \bar{U} & \bar{V} \\ \bar{V} & \bar{U} \end{bmatrix} \begin{bmatrix} C & \\ & C^\star \end{bmatrix}, \tag{44}$$

where the $(g-1)N_{\text{tot}} \times (g-1)N_{\text{tot}}$ matrices $D$ and $C$ are unitary and both $\bar{U}$ and $\bar{V}$ are real matrices of the following block diagonal form:

$$\bar{U} = \begin{bmatrix} 0 &&&&&&&&& \\ & \ddots &&&&&&&& \\ && 0 &&&&&&& \\ &&& u_1 & 0 &&&&& \\ &&& 0 & u_1 &&&&& \\ &&&&& \ddots &&&& \\ &&&&&& u_n & 0 && \\ &&&&&& 0 & u_n && \\ &&&&&&&& 1 & \\ &&&&&&&&& \ddots \\ &&&&&&&&&& 1 \end{bmatrix}, \tag{45}$$

$$\bar{V} = \begin{bmatrix} 1 & & & & & & & & & & \\ & \ddots & & & & & & & & & \\ & & 1 & & & & & & & & \\ & & & 0 & v_1 & & & & & & \\ & & & -v_1 & 0 & & & & & & \\ & & & & & \ddots & & & & & \\ & & & & & & 0 & v_n & & & \\ & & & & & & -v_n & 0 & & & \\ & & & & & & & & 0 & & \\ & & & & & & & & & \ddots & \\ & & & & & & & & & & 0 \end{bmatrix}. \quad (46)$$

In light of this decomposition of $T$, the transformation (37) defining the quasi-particle excitations can be considered as consisting of three parts.

1. A unitary transformation of the $c$-fermion creation and annihilation operators among themselves. This transformation does not mix creation operators with annihilation operators and defines a set of operators, which we denote by $a$, known as the canonical basis

$$\begin{bmatrix} a \\ a^\dagger \end{bmatrix} \equiv \begin{bmatrix} D^\dagger & \\ & D^T \end{bmatrix} \begin{bmatrix} c \\ c^\dagger \end{bmatrix}. \quad (47)$$

2. A Bogoliubov transformation which defines three classes of energy levels:

$$\begin{bmatrix} \alpha \\ \alpha^\dagger \end{bmatrix} \equiv \begin{bmatrix} \bar{U} & \bar{V}^T \\ \bar{V}^T & \bar{U} \end{bmatrix} \begin{bmatrix} a \\ a^\dagger \end{bmatrix}. \quad (48)$$

- Paired levels with $u_p, v_p > 0$:

$$\alpha_p^\dagger = u_p a_p^\dagger - v_p a_{\bar{p}}, \qquad \alpha_{\bar{p}}^\dagger = u_p a_{\bar{p}}^\dagger + v_p a_p, \quad (49)$$

where the pairs $(p,\ \bar{p})$ are defined by the $2\ \times\ 2$ blocks in (46).

- Occupied levels where $v_i = 1, u_i = 0$:

$$\alpha_i^\dagger = a_i, \qquad \alpha_i = a_i^\dagger. \quad (50)$$

- Empty levels where $v_m = 0, u_m = 1$:

$$\alpha_m^\dagger = a_m^\dagger, \qquad \alpha_m = a_m. \quad (51)$$

3. A unitary transformation of the $\alpha$ operators among themselves

$$\begin{bmatrix} \gamma \\ \gamma^\dagger \end{bmatrix} \equiv \begin{bmatrix} C^\dagger & \\ & C^T \end{bmatrix} \begin{bmatrix} \alpha \\ \alpha^\dagger \end{bmatrix}. \quad (52)$$

We can now express the state $|\phi\rangle$ in the canonical basis defined above as follows

$$|\phi\rangle = \mathcal{N}\prod_k \gamma_k|-\rangle = \mathcal{N}\prod_k \alpha_k|-\rangle, \quad (53)$$

$$= \mathcal{N}\prod_i \alpha_i \prod_p (\alpha_p \alpha_{\bar{p}})|-\rangle, \quad (54)$$

$$= \prod_i a_i^\dagger \prod_p (u_p + v_p a_p^\dagger a_{\bar{p}}^\dagger)|-\rangle, \quad (55)$$

where in the first line we have used the fact that the $\gamma$ operators are obtained by transforming the $\alpha$ operators among themselves. In the second line we have split the product of $\alpha$ operators into a product over the occupied levels and paired levels defined by the Bogoliubov transformation, the empty levels are omitted by definition of $|\phi\rangle$. In the last line we use (49) and (50) to write $|\phi\rangle$ in the canonical basis. The normalizing constant is canceled by a product of $v$ coefficients over the paired levels $\prod_p\ v_p$.

From (55) it is easy to see that $|\phi\rangle$ is a superposition of states with an even or odd particle number depending on the parity of occupied levels $\alpha_i^\dagger$. Since the occupied levels in the canonical basis have coefficients $v_i = 1$, $u_i = 0$, the number of occupied states is given by the number zeros on the diagonal of $\bar{U}$. As implied by the Bloch–Messiah theorem, we can compute $\bar{U}$ by calculating the singular value decomposition of $U$:

$$U = D\bar{U}C. \tag{56}$$

Hence, we can calculate the parity of occupied states by counting the parity of zero singular values of $U$ and in this sense we can decide which formula in (40) is the appropriate one to use when calculating the ground state energy of the system.

As well as (55), the matrix $T$ also enables us to describe $|\phi\rangle$ via its density operator and pairing tensor. Using the matrices $U$ and $V$ we can construct the generalized density operator,

$$R = \begin{bmatrix} \rho & k \\ -k^\star & 1-\rho^\star \end{bmatrix} \equiv \begin{bmatrix} V^\star \\ U^\star \end{bmatrix} [V^T \;\; U^T] \tag{57}$$

$$= \begin{bmatrix} V^\star V^T & V^\star U^T \\ U^\star V^T & U^\star U^T \end{bmatrix}. \tag{58}$$

The matrices $\rho$ and $k$ (known as the single particle density operator and the abnormal density operator or pairing tensor respectively) determine the state $|\phi\rangle$ uniquely [15]. In this sense $|\phi\rangle$ is represented by the matrix $T$.

Given that $T$ represents $|\phi\rangle$ as described above, we can construct a matrix $\widetilde{T}$ that represents the state $\gamma_1^\dagger|\phi\rangle$ by exchanging the first column of $U$ and $V$ with the first column of $V^\star$ and $U^\star$ respectively [15]. In general, we can represent any many-quasi-particle state $\gamma_{i_1}^\dagger \cdots \gamma_{i_n}^\dagger|\phi\rangle$ by interchanging columns $i_1 \cdots i_n$ of $U$ and $V$ with the corresponding columns of $V^\star$ and $U^\star$ respectively.

## 6. Ground state degeneracy

We can use the formalism developed so far to numerically calculate the ground state degeneracy of the model on a surface of arbitrary genus $g$, given enough computational resources to handle $(g-1) \times N_{\text{tot}}$ sites. As described in the last section, we can use this formalism to restrict attention to a particular common eigenspace of the vortex and loop operators and obtain an effective Hamiltonian for the fermions within that subspace. The unique ground state of this effective Hamiltonian can be found using the BdG formalism and we call it the fermionic ground state for the associated subspace.

According to the generalized flux phase conjecture the ground state of the model is in the common eigenspace of the vortex operators where all the corresponding eigenvalues are 1 (the vortex free sector of the Hilbert space). This was verified by Lieb [17] for lattices with certain periodicity. As the defect plaquettes break the translational symmetry of the lattices we are considering, the same periodicity cannot be realized and so the proof of Lieb is not applicable for our purposes. However, calculating the ground state energy of the model with different vortex configurations shows the introduction of vortices in or around the defect plaquettes increases the energy of the model. We thus assume the conjecture to be true for the lattices we are considering. We would like to emphasize that this assumption is not true for general lattices. It is possible to induce excitations in the model by including defects of a different type from the defects we have introduced here. Such defects were explored in [10] were it was found excitations induced in this way may be annihilated by vortex excitations.

Within the vortex free sector, we call the different common eigen-subspaces of the loop operators homology sectors. The degeneracy arises from the different homology sectors having fermionic ground states with the same energy. Hence, to calculate the ground state degeneracy of the system we need to calculate the fermionic ground state energy in each homology sector of the vortex free sector and see which ones have ground states with the same energy.

We considered the case of a genus $g = 2$ lattice, with both $N_a$ and $N_c$ being even numbers, and applied the analysis described above. We found that the system has a ground state degeneracy of 16 in the Abelian phase and 10 in the non-Abelian phase. In figure 10(a), we plot the difference in energy between the fermionic ground states in each of the 16 homology sectors and the homology sector with the lowest ground state energy as a function of $J = J_x = J_y$ while we fix $J_z = 1$, $\kappa = 0.2$ and $N_a = N_b = N_c = 4$. We see that the system with even dimensions $N_a$ and $N_c$ in the Abelian phase ($J < 0.5$) all 16 homology sectors are degenerate but as the system approaches the phase transition at $J = 0.5$ these sectors split with 6 of them becoming excited states in the non-Abelian phase ($J > 0.5$) while the other 10 sectors form the degenerate ground state.

When both $N_a$ and $N_c$ are odd numbers we find a eight-fold degeneracy of the ground state n the Abelian phase and ten in the non-Abelian phase. The different degeneracies we find in the Abelian phase are a result of (20). For lattices where both $N_a$ and $N_c$ are even, the parity of fermions in the ground state is the same for each homology sector. However, for lattices where either $N_a$ or $N_c$ are odd, the parity of fermions in the ground state is odd in half of the homology sectors and even in the other half. This leads to a splitting in the energy between fermionic ground states in half of the homology sectors from the other half resulting in the degree of degeneracy $d = 8$. In figure 10(b), we see that the system with odd dimensions $N_a$ and $N_c$ has half of its homology sectors forming the ground state in the Abelian phase while the other half are excited states. As the system approaches the phase transition, the sectors forming the ground state begin to split with two of them becoming excited in the non-Abelian phase while four of the excited sectors drop in energy to join the remaining six non-excited sectors to form the ten-fold degenerate ground state in the non-Abelian phase.

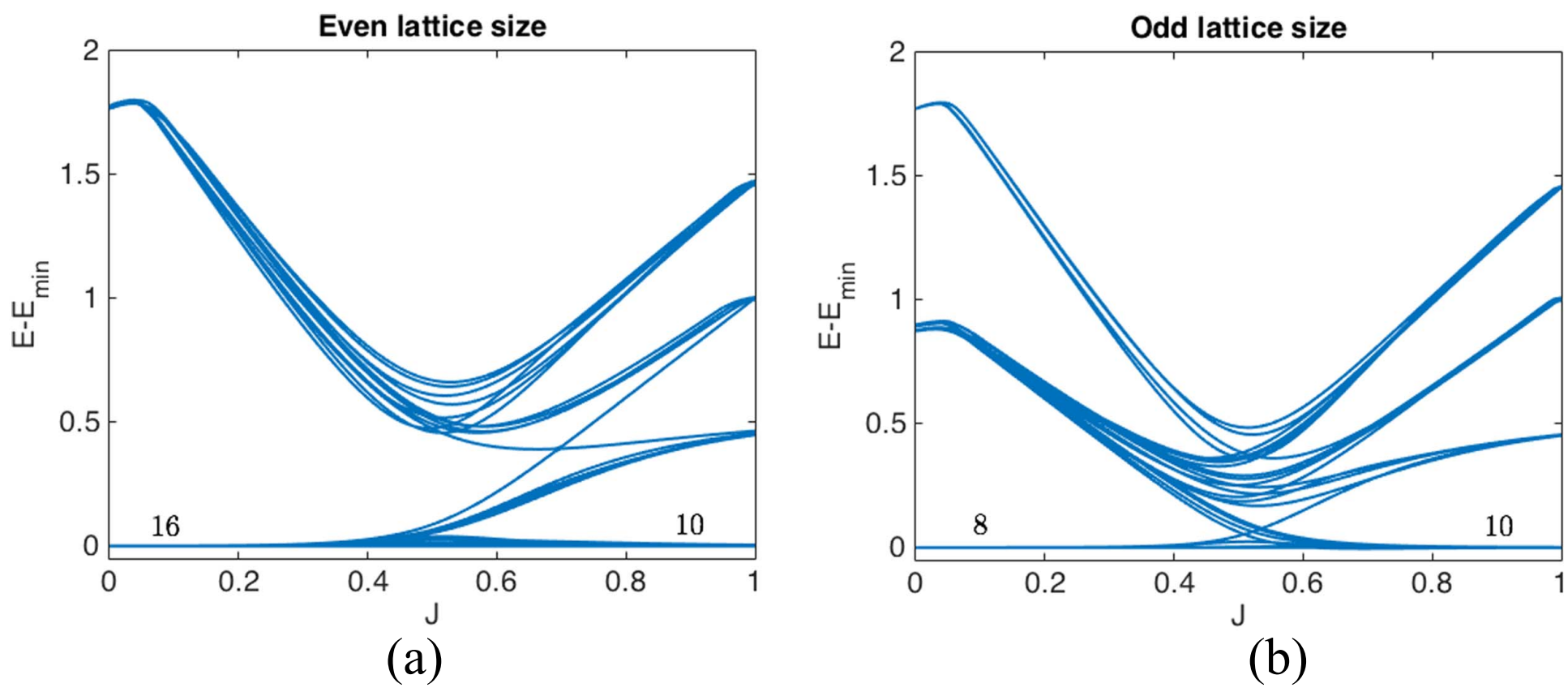

**Figure 10.** In (a) we show the difference between $E_{\min}$ and the energy $E$ of fermionic ground states and first excited states in each homology sector as a function of $J_x = J_y = J$ for $N_a = N_b = N_c = 4$ and $\kappa = 0.2$ on a genus 2 lattice. In (b) the same energy difference is plotted for $N_a = N_b = N_c = 5$. The number of degenerate ground states is included just above the lowest curves in both the Abelian and non-Abelian phases.

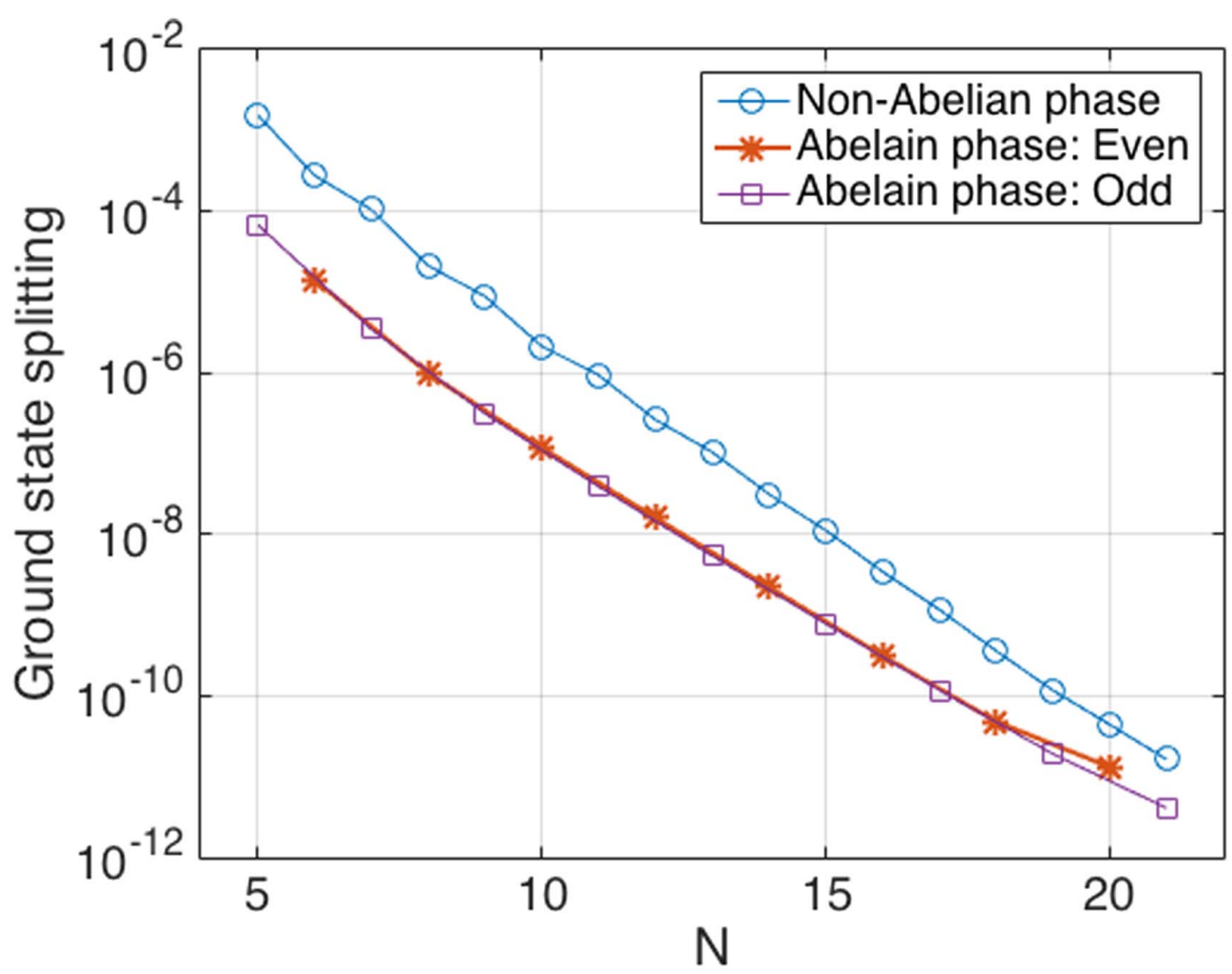

**Figure 11.** The splitting in energy between the degenerate homology sectors, measured by the difference between the sector with the highest energy and the sector with lowest energy, vanishes exponentially as the system size $N = N_a = N_c$ increases. The largest system size corresponds to over 5000 spins of the original honeycomb lattice; beyond that the numerical precision starts competing with the ground state splitting.

Due to finite size effects, there is a small splitting in the energy between the degenerate homology sectors that form the ground state. We expect this spitting to vanish in the thermodynamic limit. We measure this splitting by the difference in energy between the sector with the highest energy and the sector with the lowest energy. In figure 11 we plot the splitting between the the degenerate states as a function of $N = N_a = N_b = N_c$ for the two Abelian cases (even and odd sizes) and the non-Abelian case. As shown in the figure, we find the splitting between the sectors forming the ground state approaches zero exponentially as $N$ grows. This calculation was done with $\kappa = 0.2$ in each case and with $J = 0.1$ for both of the Abelian cases and $J = 1$ for the non-Abelian case.

**Table 1.** The ground state degeneracy for lattices of genus $g$.

| Phase | $g = 2$ | $g = 3$ | $g = 4$ | $g = 5$ | $g = 6$ |
|---|---|---|---|---|---|
| Abelian: odd | 8 | 32 | 128 | 512 | 2048 |
| Abelian: even | 16 | 64 | 256 | 1024 | 4096 |
| non-Abelian | 10 | 36 | 136 | 528 | 2080 |

We used this method to calculate the degeneracy of the system on lattices with genus $g = 2, 3, 4, 5$ and 6 in both the Abelian (for even and odd sizes) and non-Abelian phases. We have summarized the results in table 1. Kitaev showed using perturbation theory that the honeycomb model in the Abelian phase is equivalent to the toric code which can be shown to have a ground state degeneracy of $4^g$. This agrees with our results for systems with even $N_a$ and $N_c$. For systems where either $N_a$ or $N_c$ are odd we find the degeneracy is exactly half of $4^g$. This can be attributed to the fact that the equivalent toric code in this case has a line defect in it like the one discussed in [18]. Our results for the non-Abelian phase agree with a formula discussed by Oshikawa *et al* [19] who showed that the bosonic Pfaffian state, which belongs to the same universality class, has a ground state degeneracy $2^{g-1}(2^g + 1)$ given by the number of even spin structures on a surface of genus $g$ [20].

## 7. Conclusion

In summary, we realized the Kitaev honeycomb model on surfaces with genus $g \geqslant 2$ by introducing extrinsic defects to the underlying lattice. This required a non-trivial generalization of the exact solution of the model to include extra loop symmetries associated with homologically non-trivial loops which are introduced by increasing the genus of the lattice. We also highlight the dependence of the fermion parity on both the vortex and loop symmetries of the model for various lattice dimensions. The generalized solution was then used to calculate the ground states in both the Abelian and non-Abelian phases of the model. The degree of degeneracy of these ground states in both topological phases are in accord with available theoretical predictions based on topological quantum field theory.

Our work provides a direct realization of two distinct topological quantum field theories, specifically the Abelian doubled-$\mathbb{Z}_2$ and non-Abelian Ising theory, on closed surfaces of higher genus. As such it provides a solid basis for further investigation of the model on various manifolds, including also manifolds with boundaries which would extend previous studies of the Kitaev model [21]. Recent works on time-dependent simulation of creation and annihilation of vortex-like excitation on defects in the Kitaev model on torus [10] suggest the possibility of a dynamical process where creation and annihilation of extrinsic defects would result in dynamical change of the model genus and thus its topology. Interestingly this incarnation of topological field theory would be close to its axiomatic definition as a modular functor from a monoidal category of cobordisms to that of vector spaces [4, 5].

## Acknowledgments

We thank Graham Kells and Paul Watts for inspiring discussions and critical comments on the work. We wish to acknowledge the DJEI/DES/SFI/HEA Irish Center for High-End Computing (ICHEC) for the provision of computational facilities and support. Our research is supported by the Science Foundation Ireland under Principal Investigator Award No. 10/IN.1/I3013.